\documentclass[twocolumn,showpacs,prb]{revtex4}
\usepackage{amsmath}
\usepackage{graphicx}
\usepackage{dcolumn}
\usepackage{bm}

\begin{document}
% ------------------------------------------------------------------------------------------------------------ %

\title{Transport properties of n-type ultrananocrystalline diamond films}
\author{I.~S.~Beloborodov$^{1,2}$}
\author{P.~Zapol$^{1,3,4}$}
\author{D.~M.~Gruen$^{1}$}
\author{L.~A.~Curtiss$^{1,3,4}$}
\address{$^{1}$Materials Science
Division, Argonne National Laboratory, Argonne, Illinois 60439, USA \\
$^{2}$Department of Physics, University of Chicago, Chicago,
Illinois 60637, USA
\\$^{3}$Center for Nanoscale Materials, Argonne National
Laboratory, Argonne, Illinois 60439, USA \\
$^{4}$ Chemistry Division, Argonne National Laboratory, Argonne,
Illinois 60439, USA }

\date{\today}
\pacs{81.05.Uw, 73.22-f, 73.63.Bd, 72.10.-d}

\begin{abstract}
We investigate transport properties of ultrananocrystalline diamond films for a broad range of
temperatures. Addition of nitrogen during plasma-assisted growth increases the conductivity of
ultrananocrystalline diamond films by several orders of magnitude. We show that films produced at
low concentration of nitrogen in the plasma are very resistive and electron transport occurs via a
variable range hopping mechanism while in films produced at high nitrogen concentration the
electron states become delocalized and the transport properties of ultrananocrystalline diamond
films can be described using the Boltzmann formalism. We discuss the critical concentration of
carriers at which the metal to insulator transition in ultrananocrystalline diamond films occurs
and compare our results with available experimental data.
\end{abstract}

\maketitle

\section{Introduction}

Ultrananocrystalline diamond (UNCD) films have many potential applications in high-temperature
electronics because of the highest n-type conductivity and carrier concentration demonstrated for
phase-pure diamond thin films.~\cite{Gruen99,Bhattacharyya01,achatz06} Experimentally, n-type
conductivity of UNCD can be tuned within several orders of magnitude by nitrogen addition to the
feed gas during plasma-assisted growth. Without the nitrogen addition to the plasma, the resulting
films are quite resistive, but addition of more than about 8\% of nitrogen leads to a metallic
behavior. Interestingly, the grain size increases from 4 nm to 16 nm, as well as the grain boundary
width from about 0.5 nm to 2.2 nm with N concentration increase from 0\% to 20\%, but the overall
fraction of diamond phase remains consistently very high.~\cite{birrell02} At the same time,
nitrogen concentration in the films remains nearly the same, suggesting that change in the carrier
concentration due to higher number of nitrogen donors cannot be invoked to explain the increase of
conductivity. It was shown by careful Hall measurements~\cite{Williams04} that the conductivity is
n-type, and that carrier concentrations in the films increase much more than carrier mobilities
with increase in feed gas nitrogen concentration. Therefore, these studies suggested that there is
a metal-insulator transition in UNCD upon nitrogen addition, and that this transition is triggered
by changes in UNCD grain and grain boundary sizes.

Theoretical tight-binding  studies~\cite{Cleri99} of UNCD without nitrogen suggested a hopping
transport mechanism, where the donor sites are located in the grain boundaries and the bulk of the
grains is electrically inactive. Here high-angle high-energy grain boundaries were considered to be
a model of general grain boundaries of UNCD. The donor states were calculated to be carbon
$\pi$-states and dangling bond states in the grain boundaries. The hopping conductivity of UNCD was
estimated to be $10^{-6} (\Omega cm)^{-1}$, 3 to 7 orders of magnitude higher than in CVD diamond.
\begin{figure}[t]
\includegraphics[width=0.6\linewidth]{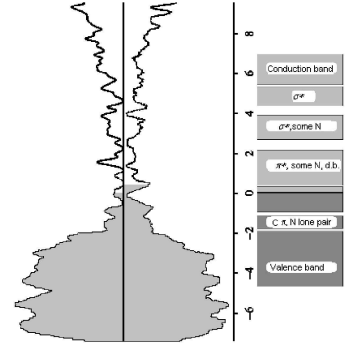}
\caption{Calculated DOS for diamond with $\Sigma13$ grain boundary and without (left) or with
(right) nitrogen (relative to the Fermi level of undoped system). Occupied states are shaded.
Sketch of electron levels in the presence of nitrogen is shown on the right. The data are taken
from Ref.~\onlinecite{Zapol2001}.}\label{figdos}
\end{figure}

Self-consistent charge density functional based tight-binding studies~\cite{Zapol2001} of nitrogen
sites in the bulk and in the grain boundaries found energies, geometries and electronic structure
of nitrogen in UNCD at low and high nitrogen concentrations. First, it was found that nitrogen is
energetically preferred in the grain boundary relative to the bulk diamond sites in the grains.
Second, nitrogen was found not to be a shallow donor, being preferentially in non-doping
configurations. Third, electrons from high-lying donor states formed upon nitrogen addition
transfer down in energy to the Fermi level, located in the $\pi$-state region.  This causes an
upward shift of the Fermi level with increase in nitrogen concentration. Calculated DOS for diamond
grain boundaries with and without nitrogen are shown in fig.~\ref{figdos} along with a scheme of
electron levels in the latter case. In addition, the connectivity of the $\pi$-bonds in the grain
boundaries improves upon nitrogen addition. This Fermi level shift and improved connectivity were
proposed to be responsible for increase in hopping conductivity.
\begin{figure}[t]
\includegraphics[width=0.6\linewidth]{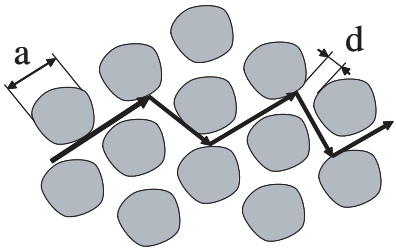}
\caption{Sketch of ultrananocrystalline diamond film with a
schematic electron path. The grey regions are insulating grains
with $a$ and $d$ being the characteristic grain size and distance
between the grains respectively.}\label{fig.grains}
\end{figure}

However, there is still no satisfactory mesoscopic description of UNCD transport properties. No
explanation was provided previously for the transition from hopping to semimetallic regime with
increase in N concentration. In this paper we consider theoretically transport properties of UNCD
films in the temperature range
\begin{equation}
10 K < T < 700K,
\end{equation}
and investigate UNCD conductivity within approximations developed
for an array of metallic and semiconducting
grains~\cite{beloborodov05,beloborodov06} in both Ohmic and
non-Ohmic regimes.

\section{Insulating regime}
\label{insulating}

For low concentrations of nitrogen ($< 1\%$ in the feed gas), the resistance of UNCD films per
square is $R \sim 10 - 100 \, M\Omega \gg R_c$, where $R_c = 6.25 k\Omega$ is the quantum
resistance. In this case, UNCD films are insulators.

The characteristic grain sizes, see Fig.~\ref{fig.grains}, of UNCD
films are in the range
\begin{equation}
4 nm < a < 16 nm
\end{equation}
and characteristic width of the grain boundaries is
\begin{equation}
0.5 nm < d < 2.2 nm.
\end{equation}
The charging energies $E_c = e^2/\kappa d$, with $\kappa$ being the dielectric constant,
corresponding to these grain boundary widths are
\begin{equation}
\label{charging} \frac{1}{\kappa}\,\, 10^3 \, K < E_c < \frac{1}{\kappa} \,\, 4 \times 10^4 K.
\end{equation}
Assuming that the dielectric constant for UNCD films is $\kappa
\sim 3-6$ from Eq.~(\ref{charging}) follows that the temperature
$T$ satisfies the inequality
\begin{equation}
T < E_c.
\end{equation}
In this temperature range the activation conductivity,
\begin{equation}
\label{activation} \sigma \sim \exp(-E_c/T)
\end{equation}
(tunneling between the nearest neighbors), is suppressed and
electron transport occurs through tunneling via several grains.

Below we assume that the concentration of donors in the grain
boundaries, $n_d$, satisfies the condition
\begin{equation}
\label{concentration} n_d\, \xi^3 \ll 1,
\end{equation}
where $\xi$ is the localization length. Using the fact that
\begin{equation}
\label{d} \xi \sim d \sim 1 \, {\rm nm},
\end{equation}
we obtain the following condition for concentration of donors
\begin{equation}
n_d \ll \, 10^{21}  \, \, {\rm cm}^{-3}.
\end{equation}
We first consider phonon assisted (Ohmic) transport in UNCD films
and later discuss the non-Ohmic regime.

\subsection{Ohmic transport}

Following the Mott-Efros-Shklovskii theory~\cite{Mottbook,Shklovskiibook} we write the hopping
conductivity in the following form
\begin{equation}
\label{f2} \sigma \sim \exp\left( -\frac{r}{\xi} -
\frac{e^2}{\kappa r T} \right),
\end{equation}
where $\xi$ is the localization length ($\xi$ is of the order of a
boundary width, $d$). Equation~(\ref{f2}) is valid for $3D$
samples and thick films with the sample thickness $L$ larger than
the optimal hopping length $r_{opt}$ defined below. This is a
typical case for UNCD films, which are grown to thicknesses of
$\mu m$,~\cite{Bhattacharyya01}. Minimizing Eq.~(\ref{f2}) with
respect to the hopping distance, $r$, we obtain the optimal
hopping length
\begin{equation}
\label{r1} r_{opt}(T) = \xi  \sqrt{\frac{e^2}{\kappa \xi T}} = \xi
\sqrt{\frac{T_0}{T}},
\end{equation}
where $T_0 = e^2/\kappa \xi$ is the characteristic energy scale.
From Eq.~(\ref{r1}) follows that the hopping length $r_{opt}(T)$
increases with decreasing temperature. Using the fact that the
hopping length
\begin{equation}
r_{opt}(T) = N(T)\, a ,
\end{equation}
where $N(T)$ is the number of grains in one hop, we obtain
\begin{equation}
N(T) = \frac{\xi}{a}\, \sqrt{\frac{T_0}{T}}.
\end{equation}
For temperatures $T \leq 300 K$ and $T_0 \sim 40000 K$ one obtains
\begin{equation}
N \sim 2.
\end{equation}
With increase in temperature the number of grains $N$ in one hop will decrease, leading to the
hopping distance $r_{opt}(T)$ of the order of the grain size. Also, increase in nitrogen
concentration is shown to lead to increase in the grain size. With increase in the grain size $a$
(and proportional increase in the grain boundary width $d$) $N$ will decrease as $\sqrt{a}$.
Substituting Eq.~(\ref{r1}) into Eq.~(\ref{f2}) we finally obtain Efros-Shklovskii
type~\cite{Efros} of conductivity behavior
\begin{equation}
\label{result1} \sigma \sim \exp\left(-\sqrt{T_0/T}\right).
\end{equation}
Equation~(\ref{result1}) is valid for temperatures $T < T_0$. At
temperatures $T \sim T_0$ the hopping distance $r_{opt}(T)$
becomes of the order of a grain size and Eq.~(\ref{f2}) does not
work anymore. In this temperature range the conductivity obeys the
Arrhenius law~(\ref{activation}).

We would like to discuss the possibility of the observation of the
Mott law in UNCD films. In semiconductors, the Efros-Shklovskii
law may turn to the Mott behavior with the increase of
temperature. This happens when the typical electron energy
$\varepsilon$ involved in a hopping process becomes larger than
the width of the Coulomb gap $\Delta_c$, i.e. when it falls into
the flat region of the density of states where Mott behavior is
expected. To estimate the width of the Coulomb gap $\Delta_c$ -
one compares the ES expression for the density of states
\begin{equation}
\nu(\Delta_c) \sim (\kappa/e^2)^D |\Delta_c|^{D-1},
\end{equation}
where $D$ is the dimensionality of a sample, with the bare density
of states $\nu_0$, i.e. the DOS in the absences of the long-range
part of the Coulomb interactions. Using the condition
$\nu(\Delta_c) \sim \nu_0$ we obtain
\begin{equation}
\label{crossover}
 \Delta_c = \left( \frac{\nu_0 e^{2D}}{\tilde \kappa^D} \right) ^{1\over
{D-1}  } .
\end{equation}
Inserting the value for the bare DOS, $ \nu_0 = 1/E_c \, \xi^D$,
into Eq.~(\ref{crossover}) we finally obtain
\begin{equation}
\label{delta} \Delta_c \sim E_c.
\end{equation}
Equation~(\ref{delta}) means that there is no flat region in the density of ground states and,
thus, the Mott regime is difficult to observe in UNCD films. At first glance, this conclusion
contradicts numerically obtained results using tight-binding methods~\cite{Cleri99,Zapol2001} since
non-zero densities of states at the Fermi level was obtained as a result of these calculations.
However, there are relatively few states in the numerical studies because of the number of atom
limitations on the size of the unit cells, which leads to uncertainties in the peak positions
larger than typical values of Coulomb gap of several 10K. In addition, Coulomb gap existence is
based on excitonic effects, which are not explicitly taken into account in the tight-binding
simulations. In other words, tight-binding simulations are not designed to accurately reproduce the
functional form of DOS in the immediate vicinity (within several tens of Kelvins) of the Fermi
level.

\subsection{Non-Ohmic transport}

Here we consider non-Ohmic conductivity in the temperature range
in which the Ohmic conductivity is given by Eq.~(\ref{result1}).
In the presence of electric field, $E$, the hopping conductivity
is given by the following expression
\begin{equation}
\label{eEr} \sigma \sim  \exp\left[-\frac{r}{\xi} -
\frac{e^2}{\kappa r T} + \frac{eEr}{T}\right].
\end{equation}

\subsubsection{Weak electric field}

In a weak electric field, $eE \xi < T$, the last term in the
r.h.s. of Eq.~(\ref{eEr}) is not important and the optimal hopping
length is given by Eq.~(\ref{r1}). Using Eqs.~(\ref{eEr}) and
(\ref{r1}) one obtains for the hopping conductivity in a weak
electric field the following expression
\begin{equation}
\label{weak} \sigma \sim \exp\left(-\sqrt{T_0/T}\right)\,
\exp\left[(eE\xi/T)\, \sqrt{T_0/T}\right].
\end{equation}
From Eq.~(\ref{weak}) follows that the Coulomb term becomes of the
order of the electric field term at fields
\begin{equation}
\label{E*} E^* \sim T/e\xi.
\end{equation}

\subsubsection{Strong electric field}

For electric fields $E > E^*$, where the field $E^*$ is defined by
Eq.~(\ref{E*}), the first term in the r.h.s. of Eq.~(\ref{eEr}) is
not important. The optimal hopping length in this regime is
temperature independent and is given by the following expression
\begin{equation}
\label{E} r_{opt} (E) = \xi \sqrt{E_0/E},
\end{equation}
where $E_0 = e/\kappa \xi^2 = T_0/e\xi$ (we note that $E_0 \gg
E^*$). Using Eq.~(\ref{E}) one obtains for the hopping
conductivity in a strong electric field~\cite{Shklovskii73}
\begin{equation}
\label{strong} \sigma \sim \exp\left(-\sqrt{E_0/E}\right).
\end{equation}
We note that: i) for very strong electric fields, $E \sim E_0$,
where the field $E_0$ is defined below Eq.~(\ref{E}), the hopping
picture does not work anymore since the hopping length becomes of
the order of the localization length, $r_{opt}(E_0) \sim \xi$; ii)
The current density $j$ can be expressed in term of conductivity
as follows $j = \sigma E$, where the conductivity $\sigma$ is
given by Eqs.~(\ref{weak}) or (\ref{strong}) depending on the
strength of the electric field $E$.

\section{Metallic Regime}
\label{metallic}

The transition from the hopping regime to degenerate semiconductor regime in UNCD is associated
with a strong increase in the number of carriers, whereas the mobility of carriers is not strongly
affected. This was demonstrated in Hall measurements~\cite{Williams04}. Also, higher nitrogen
concentration is related to increase in grain size and grain boundary width. For relatively large
concentrations of nitrogen in the plasma, ($\gg 5\%$), the UNCD film resistance becomes less than
the quantum resistance, $R \sim 100 \, \Omega \ll R_c$. As a result, the electron states become
delocalized and the transport properties of UNCD films can be described using the Boltzmann
equation.

There are several scattering mechanisms which contribute to the
resistance, $R$, of a sample~\cite{Abrikosovbook}: 1) scattering
of electrons on static impurities, $R_{ei}$, 2) electron-phonon
scattering, $R_{e-ph}$, and 3) electron-electron scattering,
$R_{e-e}$. At high temperatures the interference between different
mechanisms can be neglected and the total resistance consists of a
sum of resistances
\begin{equation}
\label{resistivity} R = R_{ei} + R_{e-ph} + R_{e-e},
\end{equation}
where $R_{ei} = \alpha  \tau_{ei}^{-1}, \, \, R_{e-ph} = \alpha \tau_{e-ph}^{-1}, \, \, R_{e-e} =
\alpha \tau_{e-e}^{-1}$. Here $\alpha = m^*/(e^2 n_e)$ with $m^*$ being the effective mass of an
electron and $\tau_{ei}$, $\tau_{e-ph}$ and $\tau_{e-e}$ are the characteristic scattering times
due to scattering of electrons on static impurities, phonons and electrons respectively. We would
like to comment on the validity of Eq.~(\ref{resistivity}). Writing Eq.~(\ref{resistivity}) we
assume that different scattering mechanisms are independent. This is indeed the case at not very
low temperatures when the scattering lengths $L_{e-ph} = u/T$, with $u$ being the phonon speed and
$L_{e-e} =  v_F/T$ are smaller than the elastic mean free path, $l = v_F\tau_{ei}$. Using the
condition ${\rm max}(L_{e-ph},L_{e-e}) \simeq l$ one can find the characteristic temperature $T^*=
\tau_{ei}^{-1}$. For temperatures $T > T^*$ the interference among different mechanisms is not
important and Eq.~(\ref{resistivity}) is valid. Below we discuss different contributions to the
resistance of UNCD films.

Assuming that the mean free path of an electron is of the order of the grain boundary width $d$ for
elastic scattering time one obtains $\tau_{ei}^{-1} \sim v_F/ d$, where $v_F$ is the Fermi
velocity. The electron-phonon scattering time $\tau_{e-ph}$ is given by the following
expression~\cite{Abrikosovbook}
\begin{equation}
\label{ph} \tau_{e-ph}^{-1} = \left\{ \begin{array}{lr} T
\hspace{2.4cm}  T > \omega_D  \\
T \left(\frac{T}{\omega_D}\right)^4 \hspace{1.2cm} T < \omega_D,
\end{array}
\right.
\end{equation}
where $\omega_D$ is the Debye temperature (for UNCD films
$\omega_D \sim 1000\, K$). For the temperature range considered
here, only the $T < \omega_D$ regime is relevant. For
electron-electron scattering time $\tau_{e-e}$ one finds
$\tau_{e-e}^{-1} = T^2/\mu$, where $\mu$ is the chemical
potential. Using Eqs.~(\ref{resistivity}) and (\ref{ph}) for
temperatures $T < \omega_D$ we obtain the following expression for
the total resistance $R$ of a sample
\begin{equation}
\label{Rtotal} R = (m^*/e^2 n_e)\, [\, v_F/d + T^5/ \omega_D^4 +
T^2/\mu \,].
\end{equation}
For temperatures $T < \omega_D$ the last two terms in the r.h.s. of Eq.~(\ref{Rtotal}) are small
therefore the resistance $R$ is almost temperature independent.

\section{Crossover from insulating to metallic regime}

In this section we discuss crossover from insulating (hopping) to semimetallic regime in UNCD
films. To find a critical concentration of carriers $n_c$ at which the crossover from insulating to
semimetallic regime occurs following the Mott's approach~\cite{Mottbook}, we compare two
characteristic lengths: the screening length $r_e \approx \frac{1}{2}\frac{\xi}{(\xi^3 n)^{1/6}}$
and the localization length $\xi$. In the metallic regime the screening length $r_e$ is small and
the localization length $\xi$ is large while in the insulating regime the situation is the
opposite. The crossover region between the two regimes can be defined as $r_e \sim \xi$. Using the
above condition we obtain the following expression for the critical concentration of carriers
\begin{equation}
\label{critical} n_c \sim \frac{1}{(4\xi)^3}.
\end{equation}
Since the localization length $\xi$ is of the order of the
distance between the grains, $\xi \sim d$, see Eq.~(\ref{d}), from
Eq.~(\ref{critical}) follows that the larger the distance between
the grains the smaller the critical concentration $n_c$ is needed
to achieve the metal to insulator crossover. For a realistic $d
\sim 1 \, nm$ we obtain  the following estimate for the critical
concentration $n_c \sim (10^{19} - 10 ^{20}) \, cm^{-3}$, which
appears to be plausible.~\cite{Bhattacharyya01}

\section{Comparison with Experiment}

\begin{figure}[t]
\includegraphics[width=0.6\linewidth]{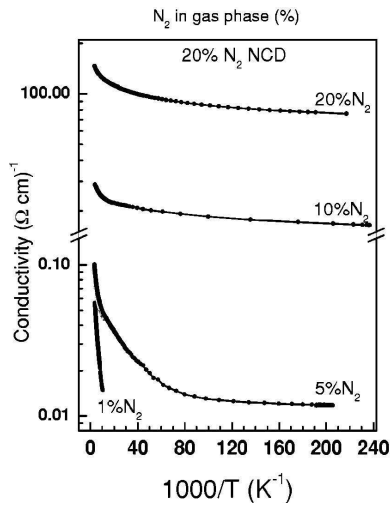}\vspace{0 cm}
\caption{Dependence of the conductivity of UNCD films on temperature. Different curves correspond
to different nitrogen concentrations in the feed gas during film growth. Reproduced from
Ref.~\onlinecite{Bhattacharyya01}}.\label{fig2}
\end{figure}

In this section we compare our results for resistivity of UNCD films with available experimental
data from Refs.~\onlinecite{Bhattacharyya01} and \onlinecite{achatz06}. These data are shown in
Fig.~\ref{fig2}. One can distinguish two qualitatively different regimes for conductivity behavior:
i) metallic, $\sigma
> 1\, \Omega^{-1} {\rm cm}^{-1}$, and ii) semiconducting, $\sigma \leq 1\, \Omega^{-1} {\rm cm}^{-1}$.

To compare our results for resistivity $R$ in the metallic regime,
Eq.~(\ref{Rtotal}), with experimental values we have taken
reasonable estimates for the critical carrier concentration of
$n_e \sim n_c \sim 10^{20} {\rm cm}^{-3}$ and $v_F$ of $10^8
\rm{cm/s}$. From Eq.~(\ref{Rtotal}), we obtain the specific
resistivity of UNCD to be $R \sim 0.1 \, \Omega \rm {cm}$. We note
that Eq.~(\ref{Rtotal}) was estimated to be valid above
temperatures $T^* = 10 - 50 \, K$.

In the semiconducting regime the conductivity behavior of UNCD films is described by
Eq.~(\ref{result1}). Since the energy scale $T_0$ in Eq.~(\ref{result1}) is inversely proportional
to the localization length $\xi$ one can see that at fixed temperature $T$ the conductivity
$\sigma$ is very sensitive to the value of localization length. This fact is in perfect agreement
with experimental data in Fig.~\ref{fig2}.

The transport properties of UNCD films have been discussed in several publications~\cite{achatz06,
bhattacharyya04}. Our treatment does not contradict the proposed DOS behavior~\cite{achatz06}. In
fact, electron densities in this work are consistent with our treatment. The main difference
between previous considerations and our work is based on our description of experimental data using
two regimes i) extended electronic states, Sec.~\ref{metallic} and ii) localized electronic states,
Sec.~\ref{insulating}. All previous interpretations of experimental data were based on electron
hopping theory only.

\section{Conclusions}

We studied transport properties of UNCD films as a function of temperature and concentration of
nitrogen during plasma-assisted growth. We show that for UNCD grown in low concentration of
nitrogen ($< 1\%$ in the feed gas)  the electron transport  occurs via a variable range hopping
mechanism while for films grown at high nitrogen concentration the electron states become
delocalized and the transport properties of ultrananocrystalline diamond films can be described
using the Boltzmann equation. Our results are in agreement with available experimental data.

\begin{acknowledgments}
This work was supported by the Division of Materials Science,
Office of Basic Energy Sciences, U.S. Department of Energy, under
Contract No. DE-AC02-06CH11357.
\end{acknowledgments}

\end{document}